\documentclass[aps, prl, showpacs, amsfonts, amsmath, amssymb, twocolumn, floatfix, superscriptaddress, 10pt]{revtex4-2}
\usepackage[final]{graphicx}
\usepackage{xcolor}
\usepackage[colorlinks=true, linkcolor=blue, citecolor=blue, urlcolor=blue]{hyperref}
\usepackage{verbatim}
\usepackage{csquotes}
\usepackage{braket}
\usepackage{mathtools}

\newcommand{\J}{\boldsymbol{\mathcal{J}}}
\newcommand{\Tr}{{\textrm{Tr}}}
\newcommand{\vct}[1]{{\bf #1}}

\renewcommand{\Im}{{\textrm{Im}}}
\renewcommand{\Re}{{\textrm{Re}}}

\DeclareSymbolFont{bbgreek}{U}{bbold}{m}{n}
\DeclareMathSymbol{\bbmu}{\mathbb}{bbgreek}{'26}
\DeclareMathSymbol{\bbeps}{\mathbb}{bbgreek}{'17}
\DeclareMathSymbol{\bbka}{\mathbb}{bbgreek}{'24}

\begin{document}


\title{Scale-dependent heat transport in dissipative media via electromagnetic fluctuations}

\author{Matthias Krüger}
\affiliation{Institute for Theoretical Physics, University of Göttingen, 37077 Göttingen, Germany}
\author{Kiryl Asheichyk}
\affiliation{Department of Theoretical Physics and Astrophysics, Belarusian State University, 5 Babruiskaya Street, 220006 Minsk, Belarus} 
\author{Mehran Kardar}
      \affiliation{Department of Physics, Massachusetts Institute of Technology, Cambridge, MA 02139, USA}
\author{Ramin Golestanian}
\affiliation{Max Planck Institute for Dynamics and Self-Organization (MPI-DS), D-37077 G\"ottingen, Germany}
\affiliation{Rudolf Peierls Centre for Theoretical Physics, University of Oxford, Oxford OX1 3PU, United Kingdom}

\date{\today}

\begin{abstract}
We develop a theory for heat transport via electromagnetic waves inside media, and use it to derive a spatially nonlocal thermal conductivity tensor, in terms of the electromagnetic Green's function and potential, for any given system. While typically negligible for optically dense bulk media, the electromagnetic component of conductivity can be significant for optically dilute media, and shows regimes of Fourier transport as well as unhindered transport. Moreover, the electromagnetic contribution is relevant even for dense media, when in presence of interfaces, as exemplified for the in-plane conductivity of a nanosheet, which shows a variety of phenomena, including absence of a Fourier regime.
\end{abstract}

\maketitle



Heat transfer through electromagnetic (EM) waves can be traced  back to Max Planck's renowned law of blackbody radiation~\cite{Planck1901}, providing the classic behavior for heat exchange by quantized traveling modes between objects separated by vacuum at appropriate large scales. Polder and van Hove introduced the concept of near-field thermal transport in 1971~\cite{Polder1971}, revealing a significant enhancement of  transfer at distances smaller than the relevant wavelengths, eliciting a mechanical response via momentum transfer \cite{Golestanian1997} and endowing vacuum with an effective friction through radiation with broken time-reversal symmetry \cite{Kardar1999}. Experimental studies have substantiated predictions of  transfer across a range of setups~\cite{Shen2009, Rousseau2009, Bimonte2017, Biehs2021}.      

For objects separated by vacuum, photons serve as the primary mechanism for energy transport. 
Within material media, other modes, e.g., kinetic, electronic, or phononic, generally provide superior routes for thermalization. Practical considerations aside, theoretical treatment of EM contributions within absorptive media is challenging due to the need for careful handling of the Poynting vector~\cite{Jackson1998}. Rigorous approaches are thus often limited to objects separated by nonabsorbing media~\cite{Muller2017}. 

Despite these limitations, investigations have been conducted into collections of nanoparticles~\cite{Ben-Abdallah2013, Ordonez-Miranda2015, Tervo2020, Luo2023} or slabs~\cite{Latella2018} separated by vacuum gaps, which ensure unimpeded radiation between separate components. These studies have revealed novel transport behavior, including \enquote{superdiffusive} and \enquote{ballistic} regimes. Similar phenomena have been studied in the context of thermally driven \cite{Najafi2004} and active \cite{Golestanian2009}  colloidal systems, where the non-local nature of the transport phenomenon  leads to the violation of fluctuation-dissipation theorem \cite{Golestanian2002}.
 
However, it has been noted that EM radiation can compete with other mechanisms even within media, for example in situations involving interfaces and surface waves~\cite{Chen2005, Ordonez-Miranda2013, Ordonez-Miranda2014, Guo2021, Ordonez-Miranda2023, Chen2010}. These predictions are based on dispersion relations for interfaces and the Boltzmann transport equation~\cite{Chen2005, Ordonez-Miranda2013, Ordonez-Miranda2014, Guo2021, Ordonez-Miranda2023}, or the Poynting vector~\cite{Joulain2008,Chen2010}. Recent experiments~\cite{Wu2020, Kim2023, Kim2023_2} have indeed demonstrated a strong increase of in-plane thermal conductivity of thin films, hinting on the strong contribution by EM surface waves. Such predictions and experimental findings underscore the need for a comprehensive and rigorous theoretical treatment of EM heat transport through materials.

\begin{figure}[t]
\begin{center}
\includegraphics[width=0.7\linewidth]{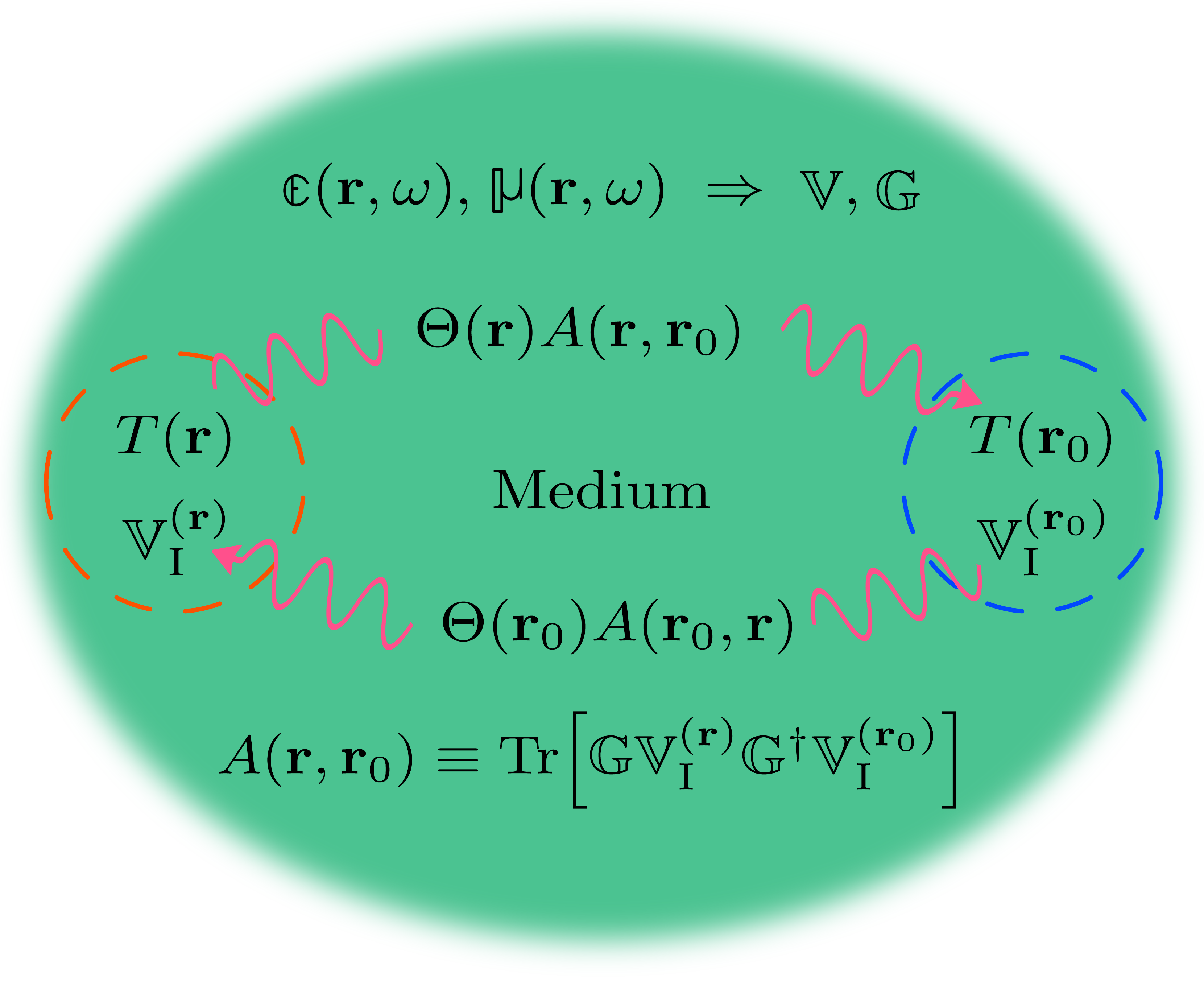}
\end{center}
\vskip-0.75cm
\caption{\label{fig:system}Medium with permittivity and permeability tensors $ \bbeps $ and $ \bbmu $, and associated electromagnetic potential $ \mathbb{V} $ and Green's tensor $ \mathbb{G} $. A temperature profile $ T(\vct{r}) $ leads to heat currents between volume elements described by Eq.~\eqref{eq:EbalanceInit}.}
\end{figure}

Here, we develop a formalism for calculating heat transport via EM fluctuations in dispersive media. The key underlying premise of the framework is the observation that the local value of the temperature field acts as a source that excites radiation fluctuations that transport energy to other regions in space (see Fig.~\ref{fig:system}), hence rendering the transport of heat through EM radiation a fundamentally nonlocal effect. By considering local Ohmic loss, we circumvent  Poynting's theorem and obtain a general expression for the nonlocal, i.e., scale-dependent thermal conductivity arising from EM waves. As applications, we investigate an optically dilute bulk medium and identify regimes of Fourier and unhindered transport. Intriguingly, radiative conduction can potentially surpass other conductive mechanisms for large decay lengths.  For a nanosheet, we uncover various phenomena originating from the slow decay of EM  modes. For distances large compared to the decay length of surface waves, the conduction kernel exhibits a power law (and not exponential) decay. This leads to the absence of a Fourier-law limit for the sheet. We  compare to the Boltzmann equation approach, delineating regimes for its (non-)validity.

Consider a medium with temperature $T(\vct{r})$ varying smoothly with position  $\vct{r}$, implying assumption of validity of a local Bose-Einstein distribution. For frequency $\omega\geq0$, the mean energy of photons emitted at $\vct{r}$ is thus~\cite{Planck1901}
\begin{equation}
\Theta(\omega,\vct{r})=\hbar\omega\left[\frac{1}{e^{\frac{\hbar\omega}{k_{\textrm{B}}T(\vct{r})}}-1}+\frac{1}{2}\right],
\label{eq:Theta}
\end{equation}
where $\hbar$ and $k_{\textrm{B}} $ are Planck's and Boltzmann's constants, respectively. The radiation sourced by the  volume element at position $\vct{r}$ is expressed in terms of the correlation of electromagnetic fields~\cite{Rytov1989} ($ c $ is vacuum speed of light), 
\begin{equation}
\mathbb{C}_{\vct{r}} \equiv \left\langle\vct{E}\otimes\vct{E}^*\right\rangle_{\omega}^{(\vct{r})} = \frac{8\pi\omega}{c^2}\Theta(\omega,\vct{r}) ~\mathbb{G}(\mathbb{V}_{\textrm{I}}\mathbb{I}_{\vct{r}})\mathbb{G}^{\dagger},
\label{eq:C}
\end{equation}
where operator notation is understood~\cite{Kruger2012} \footnote{The two spatial arguments of $\mathbb{C}$ are the outer arguments of the operator $ \mathbb{G}(\mathbb{V}_{\textrm{I}}\mathbb{I}_{\vct{r}})\mathbb{G}^{\dagger }$, and operator products involve integration over a joint coordinate and a matrix product in $ 3\times 3 $ space~\cite{Kruger2012}.}. $ \mathbb{G} $ is the system's Green's tensor, and we have introduced the EM potential $\mathbb{V}=\frac{\omega^2}{c^2}(\bbeps-\mathbb{I})+\nabla\times\left(\mathbb{I}-\bbmu^{-1}\right)\nabla\times$, with permittivity and permeability tensors $ \bbeps $ and $ \bbmu $, and identity $ \mathbb{I} $. Expressing  dissipation as $ \mathbb{V}_{\textrm{I}}\equiv\frac{\mathbb{V}-\mathbb{V}^\dagger}{2i} $ allows for optical nonreciprocity~\cite{Gelbwaser-Klimovsky2021}. We further require the EM potential to be local~\cite{Jackson1998} on scales shorter than variations of temperature, and introduce $ \mathbb{I}_{\vct{r}}=\mathcal{I}\delta^{(3)}(\vct{r}_1-\vct{r})\delta^{(3)}(\vct{r}-\vct{r}_2) $, with identity matrix $ \mathcal{I} $, to pick the source point in Eq.~\eqref{eq:C}. As $ \mathbb{V}_{\textrm{I}} $ is local, $ \mathbb{I}_{\vct{r}} $ and $\mathbb{V}_{\textrm{I}} $ commute, and $ (\mathbb{V}_{\textrm{I}}\mathbb{I}_{\vct{r}}) $ is Hermitian and nonnegative, and denoted by $ (\mathbb{V}_{\textrm{I}}\mathbb{I}_{\vct{r}}) \equiv \mathbb{V}_{\textrm{I}}^{(\vct{r})}$.

The radiation in Eq.~\eqref{eq:C} leads to absorption of energy $H_\omega$ by a volume element at $\vct{r}_0$. To avoid the Poynting vector, we directly compute the local Ohmic loss~\cite{Jackson1998}~\footnote{$ H_{\omega} $ has dimensions of energy flux density, such that integration over $ \vct{r}$, $\vct{r}_0 $, and $ \omega $ yields energy per time.},
\begin{equation}
H_{\omega}(\vct{r}\to\vct{r}_0) = \frac{1}{\pi}\Re\left\langle \vct{E}(\vct{r}_0)\cdot\vct{J}^*(\vct{r}_0) \right\rangle_{\omega}^{(\vct{r})}.
\label{eq:HPower}
\end{equation}
To proceed, we use the free Green's function $\mathbb{G}_0$ to convert between the total field and the total current~\cite{Jackson1998, Tsang2000}, $\vct{E}=4\pi i\frac{\omega}{c^2}\mathbb{G}_0\vct{J}$. Substituting $  \vct{J} $ into Eq.~\eqref{eq:HPower} and using Eq.~\eqref{eq:C} yields
\begin{equation}
H_{\omega}(\vct{r}\to\vct{r}_0) = -\frac{2}{\pi}\Theta(\vct{r})\Im\Tr\left[\mathbb{I}_{\vct{r}_0}\mathbb{G}\mathbb{V}_{\textrm{I}}^{(\vct{r})}\mathbb{G}^{\dagger}\mathbb{G}_0^{\dagger -1}\right].
\label{eq:HInterm}
\end{equation}
Using identity $\mathbb{G}^{\dagger}=\mathbb{G}_0^{\dagger}+\mathbb{G}_0^{\dagger} [\mathbb{I}-\mathbb{V}^{\dagger}\mathbb{G}_0^{\dagger}]^{-1}\mathbb{V}^{\dagger}\mathbb{G}_0^{\dagger}$ and resumming the inverse, we obtain
\begin{equation}
\mathbb{G}^{\dagger}\mathbb{G}_0^{\dagger -1} = \mathbb{I}+\mathbb{G}_0^{\dagger}\left[\mathbb{I}-\mathbb{V}^{\dagger}\mathbb{G}_0^{\dagger}\right]^{-1}\mathbb{V}^{\dagger} = \mathbb{I}+\mathbb{G}^{\dagger}\mathbb{V}^{\dagger}.
\label{eq:GG0inv}    
\end{equation}
Plugging Eq.~\eqref{eq:GG0inv} into Eq.~\eqref{eq:HInterm} gives two contributions 
\begin{equation}
\frac{H_{\omega}(\vct{r}\to\vct{r}_0)}{\frac{2}{\pi}\Theta(\vct{r})} = \Tr\left[\mathbb{G}\mathbb{V}_{\textrm{I}}^{(\vct{r})}\mathbb{G}^{\dagger}\mathbb{V}_{\textrm{I}}^{(\vct{r}_0)}-\Im\left(\mathbb{I}_{\vct{r}_0}\mathbb{G}\mathbb{V}_{\textrm{I}}^{(\vct{r})}\right)\right].
\label{eq:H}
\end{equation}
The second term 
represents re-absorption at $ \vct{r} $, and does not contribute to energy exchange, while the first one accounts for heat transport from $ \vct{r} $ to $ \vct{r}_0 $.  Notably, when $ \vct{r} $ and $ \vct{r}_0 $ are integrated over disconnected object volumes, Eq.~\eqref{eq:H} recovers the known expressions for heat transfer between separate bodies~\cite{Kruger2012, Soo2018}. Equation~\eqref{eq:H} is nonperturbative as $ \mathbb{G} $ contains $ \mathbb{V} $ to all orders, and is valid for large or small $ \mathbb{V} $. Expanding $ \mathbb{G} $ for small $ \mathbb{V} $, Eq.~\eqref{eq:H} reproduces the heat flow in systems of small particles in vacuum~\cite{Ben-Abdallah2013, Asheichyk2017}.

Balancing energy absorption and emission at $ \vct{r}_0 $ yields the divergence of energy current $ \J $; see Fig.~\ref{fig:system}. We integrate over $ \omega $ and $ \vct{r} $, to obtain ($ \nabla \equiv \nabla_{\vct{r}} $ and $ \nabla_0 \equiv \nabla_{\vct{r}_0} $)
\begin{align}  
\notag &-\nabla_0\cdot\J(\vct{r}_0)= \int_0^{\infty} \!\!\! d\omega\int \! d\vct{r} \left[H_{\omega}(\vct{r}\to\vct{r}_0)-H_{\omega}(\vct{r}_0\to\vct{r})\right]\\
& = \frac{2}{\pi}\int_0^{\infty} \!\! d\omega\int \! d^3r\left[\Theta(\omega,\vct{r})A(\vct{r},\vct{r}_0)-\Theta(\omega,\vct{r}_0)A(\vct{r}_0,\vct{r})\right].
\label{eq:EbalanceInit}
\end{align}
In the last step, we have introduced the abbreviation
\begin{equation}
A(\vct{r},\vct{r}_0) \equiv 
\Tr\left[\mathbb{G}\mathbb{V}_{\textrm{I}}^{(\vct{r})}\mathbb{G}^{\dagger}\mathbb{V}_{\textrm{I}}^{(\vct{r}_0)}\right],\label{eq:A_def}
\end{equation}
for the spectral conductivity kernel, which is nonnegative. We continue with reciprocal media, for which $ A(\vct{r},\vct{r}_0) = A(\vct{r}_0,\vct{r}) $ is symmetric, and Eq.~\eqref{eq:EbalanceInit} becomes
\begin{equation}
-\nabla_0\cdot\J(\vct{r}_0) = \frac{2}{\pi}\int_0^{\infty} \!\! d\omega\int \! d^3r\left[\Theta(\vct{r})-\Theta(\vct{r}_0)\right]A(\vct{r},\vct{r}_0). 
\label{eq:EbalanceRecip}
\end{equation}
$ A \geq 0 $ ensures that energy  flows from warmer to colder volume elements. Specializing to small variations in temperature around $ T_0 $, we linearize the Bose-Einstein weight $ \Theta(T) = \Theta(T_0) + k_{\textrm{B}}\mathcal{N}(\tilde\omega)(T(\vct{r})-T_0)+\cdots$, with dimensionless $\tilde\omega= \hbar\omega/(k_{\textrm{B}}T_0)$ and $\mathcal{N}(\tilde\omega)={\tilde\omega}^2e^{\tilde\omega}/(e^{\tilde\omega}-1)^2$. 
We get 
\begin{equation}
-\nabla_0\cdot\J(\vct{r}_0) = \frac{2}{\pi}\int \! d^3r ~\mathcal{A}(\vct{r},\vct{r}_0)\left[T(\vct{r})-T(\vct{r}_0)\right],
\label{eq:EbalanceTlin}
\end{equation}
with conductivity kernel $ \mathcal{A}(\vct{r},\vct{r}_0) \equiv k_{\textrm{B}}\int_0^{\infty} \! d\omega \mathcal{N}(\tilde\omega)A(\vct{r},\vct{r}_0) $, which, as $ \mathcal{N} \geq 0 $, is nonnegative as well. Equation~\eqref{eq:EbalanceTlin} can be turned in a gradient expansion in $ T $, by expanding $ T(\vct{r}_0) $ around $ \vct{r} $, $ T(\vct{r})-T(\vct{r}_0) = \left(1-e^{(\vct{r}_0-\vct{r})\cdot \nabla}\right)T(\vct{r}) $, so that
\begin{equation}
-\nabla_0\cdot\J(\vct{r}_0) = \frac{2}{\pi} \int \! d^3r ~\mathcal{A}(\vct{r},\vct{r}_0)\left(1-e^{(\vct{r}_0-\vct{r})\cdot \nabla}\right) T(\vct{r}).
\label{eq:EbalanceTnabla}
\end{equation}
Matching Eq.~\eqref{eq:EbalanceTnabla} with a nonlocal Fourier law,
\begin{equation}
-\nabla_0\cdot\J(\vct{r}_0) = \nabla_0\cdot\int \! d^3r ~\bbka(\vct{r}_0,\vct{r}) \cdot \nabla T(\vct{r}), 
\label{eq:FourierLaw}
\end{equation}
we  obtain the nonlocal conductivity tensor~\footnote{The Fourier transform of $\bbka(\vct{r}_0,\vct{r})$ carries units of conductivity.}
\begin{equation}
\bbka(\vct{r}_0,\vct{r}) = \frac{2}{\pi}\nabla_0^{-1}\mathcal{A}(\vct{r},\vct{r}_0)\left(1-e^{(\vct{r}_0-\vct{r})\cdot \nabla}\right)\otimes \nabla^{-1}.
\label{eq:kappa}
\end{equation}
Equation~\eqref{eq:kappa} is a main result of this work, providing the conductivity tensor from EM waves inside media, for general shapes and inhomogeneities (as encoded in  $ \mathbb{G} $).

For uniform  isotropic bulk, $ A(\vct{r},\vct{r}_0) = A(|\vct{r}-\vct{r}_0|) $, recommending  a 3D Fourier transform from $ \vct{r}-\vct{r}_0 $ to $ \vct{q} $, using $ e^{(\vct{r}_0-\vct{r})\cdot \nabla}e^{i\vct{q}\cdot  (\vct{r}_0-\vct{r})} = 1 $. $ \bbka = \mathcal{I}\kappa $ is then diagonal and depends on $ q \equiv |\vct{q}| $,
\begin{equation}
\kappa(q) = \frac{16\pi^2k_{\textrm{B}}}{q^2} \int_0^{\infty} d\omega~ \mathcal{N}(\tilde\omega)\left[A(q=0)-A(q)\right],
\label{eq:kappa_bulk}
\end{equation}
where $ A(q) = \frac{1}{2\pi^2q} \int_l^{\infty} \! dr A(r) r\sin(qr) $, with $ r \equiv |\vct{r}-\vct{r}_0| $. A lower cutoff length $ l \geq 0 $ reflects the breakdown of the continuum approach at atomistic scales. Since $ qr \geq \sin(qr) $ and  $ \frac{d}{dq} \frac{1}{q^3}\left[qr-\sin(qr)\right]\leq0 $, $ \kappa $ is positive and decays monotonically with $ q $ and  with $l$.

The Green's tensor for a nonmagnetic isotropic homogeneous medium is found from $ \mathbb{G}_0 $ by replacing the vacuum speed of light $c$ by the complex speed of light in the medium $ c/\sqrt{\varepsilon} $, with wavenumber $ k=\frac{\omega}{c}\sqrt{\varepsilon} $. As both emission and absorption are governed by $ \mathbb{V}_{\textrm{I}} = \frac{\omega^2}{c^2}\Im[\varepsilon]\mathbb{I} $, Eq.~\eqref{eq:A_def} yields (neglecting singular parts at $r=0$)~\cite{Tsang2000}
\begin{align}
\notag A(r) & = \frac{\Im[\varepsilon]^2\frac{\omega^4}{c^4}}{8\pi^2 r^2} e^{-2\Im[k]r}\biggl[1+\frac{2\Im[k]}{r|k|^2}\\
& +\frac{|k|^2+4\Im[k]^2}{|k|^4r^2}+\frac{6\Im[k]}{r^3|k|^4}+\frac{3}{|k|^4r^4}\biggr].
\label{eq:Abulk}
\end{align}
Using this in Eq.~\eqref{eq:kappa_bulk} allows to find $\kappa$ as a function of $q$. It is insightful to start by expanding the integrand of $ \kappa(q) $ for small $q$. This yields a regime of $q$-independent Fourier conductivity, involving  integrals 
\begin{equation}
I_m=\int_l^{\infty} \!\! dr ~r^m  e^{-2\Im[k]r},
\label{eq:I}
\end{equation}
where $ m \in \{2,1,0,-1,-2\} $ from the terms in Eq.~\eqref{eq:Abulk}. From small $r$ behavior, $I_m\sim l^{m+1}$, i.e., the terms $ m=-1 $ and $ m = -2 $ depend on the cutoff as $\log(l)$ and $l^{-1}$, respectively. 
Numerical evaluation for a dense medium, such as doped Si (see $ \varepsilon_{\textrm{Si}} $ below), yields $ \kappa $ negligible compared to other conduction mechanisms.

\begin{figure}[!t]
\begin{center}
\includegraphics[width=1.0\linewidth]{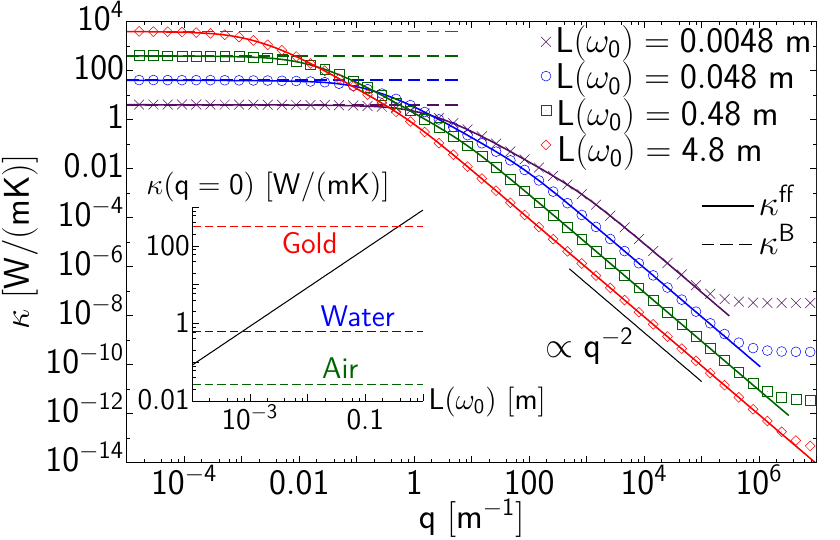}
\end{center}
\caption{\label{fig:bulk}EM heat conductivity of a dilute medium at $ T_0 = 300~\textrm{K} $ as a function of wavenumber $q$, for various lengths $L(\omega_0)$. Solid and dashed lines show analytical limits. Inset: $ \kappa(q=0) $ as a function of $ L(\omega_0) $. Dashed lines give literature values of conductivity for selected materials~\cite{Kadoya1985, Bird2002, Dean1998}.}    
\end{figure}

Notably, the behavior at large $r$, $ I_m\sim L^{m+1} $, with frequency dependent decay length $L=\Im[k]^{-1}$, implies a possibly large $\kappa$ for large $L$, which we examine with the following model of dielectric permittivity,
$\varepsilon(\omega) = 1 + \alpha\omega_{0}^2/(\omega_{0}^2-\omega^2-i
\gamma\omega)$. It contains an infrared resonance, set at $ \omega_0 = 1.26 \times 10^{14} \ \textrm{rad} \ \textrm{s}^{-1} $, and damping $ \gamma = \omega_0/10 $. For $ \alpha \ll 1 $, absorption is small, and the decay length   $L(\omega_0)\approx \frac{2c\gamma}{\alpha\omega_0^2} $, diverges as $ \alpha \to 0 $. The main panel of Fig.~\ref{fig:bulk} shows the numerically obtained conductivity as a function of $ q $ for $ \alpha = 10^{-n} $ with $ n \in \{4,5,6,7\} $, corresponding to $ L(\omega_0) = 10^{n} \times 0.48 \ \mu\textrm{m} $, i.e, $ L $  ranges between $ 4.8 \ \textrm{mm} $ and $ 4.8 \ \textrm{m} $ in the graph, as labeled.

The data in Fig.~\ref{fig:bulk} are dominated by the contribution of the far-field term in Eq.~\eqref{eq:Abulk}, which is shown as solid lines~\footnote{The far-field term can be integrated to closed form: $\kappa^{\textrm{ff}} = \frac{k_{\textrm{B}}}{\pi^2}\int_0^\infty \!\! d\omega \frac{\mathcal{N}(\tilde\omega)\Im[\varepsilon]^2\omega^4}{c^4 q^3} \left[\frac{q}{2\Im[k]}-\arctan\left(\frac{q}{2\Im[k]}\right)\right]$}. For $ q \ll \Im[k]$, the aforementioned Fourier regime of $q$-independent conductivity is approached~\footnote{Equation~\eqref{eq:kappa_q0} follows from the expansion of the far-field term contribution [i.e., evaluating  $m=2$ in  Eq.~\eqref{eq:I}] for small $ \Im[\varepsilon] $, with $ \Im[k]= \frac{\omega}{c}\frac{\Im[\varepsilon]}{2\sqrt{\Re[\varepsilon]}}+\mathcal{O}(\Im[\varepsilon]^2) $. Equivalently, it is a small $ q $ and small $ \Im[\varepsilon] $ expansion of $ \kappa^{\textrm{ff}} $ in~\cite{Note4}.}, 
\begin{equation}
\kappa^{\textrm{B}} = \frac{k_{\textrm{B}}}{3\pi^2 } \int_0^{\infty} \!\! d\omega ~\mathcal{N}(\tilde\omega) \frac{\Re[\varepsilon]^{\frac{3}{2}} \frac{\omega}{c}}{\Im[\varepsilon]},
\label{eq:kappa_q0}
\end{equation}
which agrees exactly with the conductivity starting from the Boltzmann equation~\cite{Simkin2000, Tritt2004}, by identifying group velocity $ c/|\sqrt{\varepsilon}| $ and mean free path $ L/2 $. Equation~\eqref{eq:kappa_q0} shows that the small $ q $ limit grows as $ \Im[\varepsilon]^{-1}\sim L $, diverging for the limit of empty vacuum. This is surprising as both emission and absorption vanish in this limit, yielding a factor of $ 1/L^2 $. This is compensated by $I_2\sim L^3$ in Eq.~\eqref{eq:I}. The divergence of $\kappa$ as $ L \to \infty $ is reminiscent of Olbers' paradox~\cite{Wesson1991}.

Expanding the far-field term in  Eq.~\eqref{eq:Abulk} for $ q \gg \Im[k] $ yields a regime of $ \kappa\sim q^{-2} $~\footnote{In this regime, $\kappa^{\textrm{ff}} = \frac{k_{\textrm{B}}}{\pi^2q^2}\int_0^\infty \!\! d\omega \mathcal{N}(\tilde\omega)\Im[\varepsilon]\sqrt{\Re[\varepsilon]}\frac{\omega^3}{c^3}$, following from large $ q $ and small $ \Im[\varepsilon] $ limit of $ \kappa^{\textrm{ff}} $ in~\cite{Note4}.}, in agreement with the numerical data of Fig.~\ref{fig:bulk}. Such highly nonlocal conductivity (compare also Ref.~\cite{Ben-Abdallah2013}) is a manifestation of unhindered transport of photons, and was to our knowledge not identified in previous studies of photonic transport. The conductivity from the far-field term vanishes for large $ q $, and is at $ q \approx \sqrt{lL}|k|^2 $ superseded by the cutoff-dependent near-field contribution due the last term in Eq.~\eqref{eq:Abulk}, leading to the other plateau in Fig.~\ref{fig:bulk}. This plateau follows $\sim l^{-1}$, and we use $l = 0.1 \ \textrm{nm} $ in the graph.

The inset of Fig.~\ref{fig:bulk} shows the small $ q $ limit as a function of $ L $, reaffirming the linear (cutoff-independent) growth with $ L$ of Eq.~\eqref{eq:kappa_q0}, i.e., $ \kappa/L \approx 840 \ \textrm{W}/(\textrm{m}^2\textrm{K}) $ for the chosen parameters. The inset also indicates literature conductivity values of a variety of materials for comparison, illustrating the relatively large predicted $\kappa$ of optically dilute media. The decay length of air may be estimated to be around $ 5 $ meters~\cite{Wei2018}, and for such a large decay length, the regime of unhindered transport  spans  several orders of magnitude. The predicted radiative conductivity for small $ q $ appears quite large compared to kinetic mechanisms. It is a puzzle to reconcile this observation in lieu of other mechanisms of heat transport in air.

\begin{figure}[!t]
\begin{center}
\includegraphics[width=0.98\linewidth]{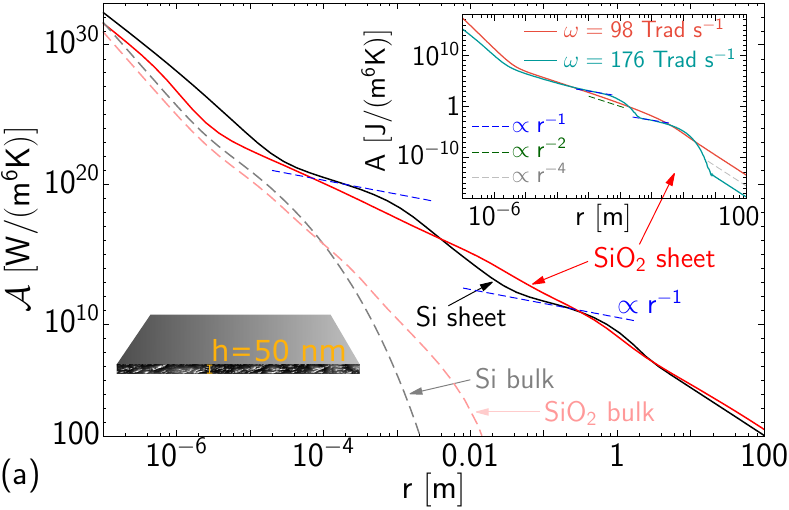}\\
\includegraphics[width=0.98\linewidth]{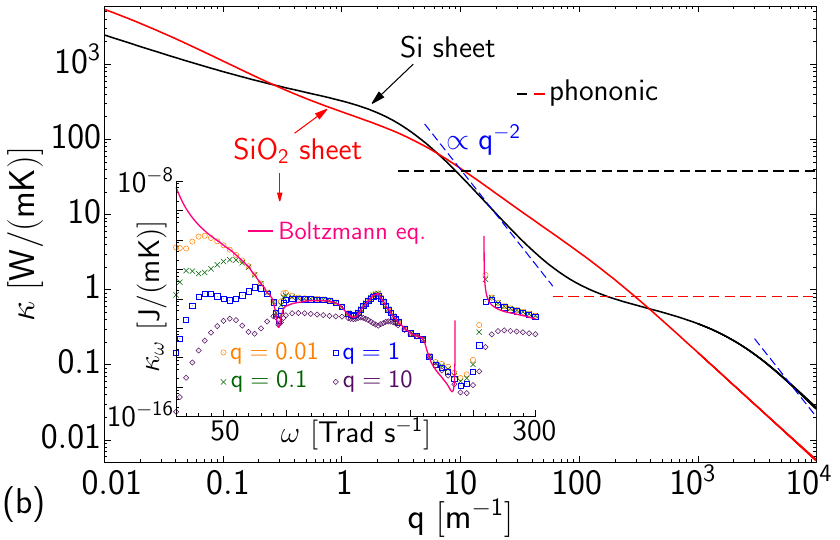}
\end{center}
\caption{\label{fig:kappa_sheet}(a) Conductivity kernel ($ T_0 = 300 \ \textrm{K} $) of a nanosheet (doped Si and $ \textrm{SiO}_2 $, $ h = 50 \ \textrm{nm} $), as a function of in-plane distance $ r $. Dashed lines indicate the bulk results. Inset: Spectral kernel for two distinct frequencies for SiO$_2$. (b) Conductivity as a function of  $ q $.  
Dashed lines give the phononic contributions~\cite{Liu2006, Zhu2018}. Inset compares the spectra for different $ q $ to the one obtained from the Boltzmann equation~\cite{Chen2005,Ordonez-Miranda2013}.}
\end{figure}

While optically dense bulk media show a negligible EM conductivity, photons can strongly contribute when considering slabs or interfaces~\cite{Chen2005, Ordonez-Miranda2013, Ordonez-Miranda2014, Guo2021, Ordonez-Miranda2023, Chen2010, Wu2020, Kim2023, Kim2023_2}. To investigate this, we consider a sheet filling the space $-h\leq z \leq 0$, with vacuum outside (inset  of Fig.~\ref{fig:kappa_sheet}). To simplify, let $ h \ll L $ for all $\omega$, so that, for in-plane distance $ r \gg h $, we may assume $ A(\vct{r},\vct{r}_0) = A(r) $ for both points inside the sheet. $ A=0 $ otherwise, because $\mathbb{V}_I=0$ in vacuum, see Eq.~\eqref{eq:A_def}. Assuming uniform temperature along $ z $, integration over $ z $ of Eq.~\eqref{eq:kappa} yields a prefactor $ h $. The parallel component of conductivity is
\begin{equation}
\kappa(q) = h\frac{8\pi k_{\textrm{B}}}{q^2} \int_0^{\infty} \!\! d\omega ~\mathcal{N}(\tilde\omega)\left[A(q=0)-A(q)\right]
\label{eq:kappa_sheet},
\end{equation}
with $ A(q) = \frac{1}{2\pi}\int_l^{\infty}drA(r)rJ_0(qr) $ the 2D Fourier transform of $ A(r) $, with $ J_0 $ the Bessel function of order zero. As in bulk, $ \kappa $ in Eq.~\eqref{eq:kappa_sheet} is a nonnegative monotonically decaying function of $ l $.

We numerically compute $ A(r) $ from the dyadic Green's function of the sheet, using a plane wave decomposition as for a planar cavity~\cite{Kim2023, Asheichyk2018}. It encodes a variety of nontrivial features, which we investigate for doped silicon and amorphous $ \textrm{SiO}_2 $, ubiquitous in microelectronics. For Si, we use the Drude model~\cite{Marquier2004, Duraffourg2006, Basu2009, Ying2020, Sze1968}, $\varepsilon_{\textrm{Si}}(\omega)=\varepsilon_{\infty}-\omega_{\textrm{p}}^2/[\omega(\omega+i\gamma)]$, with $ \varepsilon_{\infty}=11.7 $, $ \omega_{\textrm{p}} = 3.06\times 10^{14} \ \textrm{rad} \ \textrm{s}^{-1} $, and $ \gamma = 8.29\times 10^{13} \ \textrm{rad} \ \textrm{s}^{-1} $ for the chosen p-type doping concentration of $ 10^{19} \ \textrm{cm}^{-1} $; the minimal decay length is $ L \approx 2 \ \mu\textrm{m} $. For $ \textrm{SiO}_2 $, the data for $ \omega \in [12, \ 301] \ \textrm{Trad} \ \textrm{s}^{-1} $ is taken from Ref.~\cite{Palik1985}; the minimal decay length is $ L \approx 0.5 \ \mu\textrm{m} $.

Figure~\ref{fig:kappa_sheet}(a) depicts $ \mathcal{A}(r) $ for  $ h=50$ nm, covering  orders of magnitude in $ r $, to span the features on various scales.  While the bulk curves (dashed lines) decay exponentially on the scale of $ L $, the sheet solutions decay slowly. Especially Si shows two distinct plateaus with approximate scalings $ \mathcal{A} \sim r^{-1} $. We interpret these as manifestations of the two traveling surface modes, spreading as $ r^{-1} $ in 2D space, with decay lengths of roughly a millimeter and a meter. Notably,  at distances large compared to that decay length, $ \mathcal{A} $  does not decay exponentially, but with a power law. We numerically found that the spectral $ A(r,\omega) $ ultimately, for large $r$, scales as $r^{-4}$ for any $\varepsilon(\omega)$ investigated \footnote{This power law was also found for points in vacuum near a surface \cite{Messina18}.}, while, for the range shown, the integrated curves in the graph show  $\sim r^{-3.5}$.  This intriguing behavior indicates absence of an ultimate  decay length. 

For $ \textrm{SiO}_2 $, some frequencies show pronounced  plateaus in $A(r,\omega)$, e.g.  $\omega=176 \ \textrm{Trad} \ \textrm{s}^{-1} $, while they are absent for others, e.g.,  $\omega=98 \ \textrm{Trad} \ \textrm{s}^{-1} $ [inset of Fig.~\ref{fig:kappa_sheet}(a)]. Thus, the main graph of Fig.~\ref{fig:kappa_sheet}(a) shows no distinct plateaus.

Figure~\ref{fig:kappa_sheet}(b) shows the resulting conductivity $ \kappa(q) $ using cutoff $l\in [2h,10h]$ large compared to atomistic scales \cite{Qiu_2015} \footnote{For the shown range of $ q $, the results have no significant dependence on cutoff $ l \in [2h,10h] $.}. For large $ q $ the conductivity is  below the phononic contribution (shown by dashed lines), and strongly increases with  decreasing $ q $, directly related to $ \mathcal{A}(r) $:  The regimes of surface waves of $ \mathcal{A} \sim r^{-1} $ cause steep increases with nearly $ \kappa\sim q^{-2} $, i.e., nearly unhindered transport. 
For the two materials,  the EM contribution becomes comparable to the phononic one at scales of around 10 cm and a few mm, respectively. The power law seen in Fig.~\ref{fig:kappa_sheet}(a), $ \mathcal{A} \sim r^{-3.5} $, translates to $ \kappa\sim q^{-0.5} $ for small $ q $. (The mentioned behavior of $A(r,\omega)\sim r^{-4}$ yields $\kappa_{\omega}\sim\log(q)$ for small $q$.) Thus, in contrast to bulk, no $ q $-independent plateau of Fourier conduction seems to exist.  

It is thus interesting to compare to the Boltzmann equation~\cite{Chen2005, Ordonez-Miranda2013}. The   inset of Fig.~\ref{fig:kappa_sheet}(b) shows the spectral conductivity for SiO$_2$,  for various $q$. Astonishingly, the Boltzmann equation is very accurate for a range of frequencies [where plateaus are seen in $A(r,\omega)$ in Fig.~\ref{fig:kappa_sheet}(a)]. For other frequencies (e.g., $\omega=98$ Trad s$^{-1}$), it deviates notably~\footnote{For $\omega$ around $ 250 \ \textrm{Trad} \ \textrm{s}^{-1} $, the Boltzmann approach, using thin film dispersion relations \cite{Ordonez-Miranda2013}, yields negative $ \kappa_{\omega} $.}. Naive use of the Boltzmann equation yields a divergence for $\omega\to0$ seen in the graph; we interpret this to be due to absence of the Fourier regime: For a given  $q$, our data and the Boltzmann equation agree down to a certain $\omega$ (potentially corresponding to a decay length $\sim q^{-1}$ of surface modes), below which they disagree. Our $q$-dependent theory thus solidly justifies frequency cutoffs or effective propagation lengths for the Boltzmann approach \cite{Ordonez-Miranda2023, Wu2020,Kim2023}. 

While we presented applications to special cases, the formalism presented here is general, and combined with the various numerical schemes for computation of Green's functions \cite{Rodriguez12}, can deal with myriad setups, including, e.g., media in cavities \cite{Asheichyk2023}, at interfaces \cite{Ordonez-Miranda2013}, outside or inside cylinders \cite{Asheichyk2022, Asheichyk2023} or curved interfaces. Such a formalism is needed to quantify radiative effects within media and devices at nano- and microscale. Future work can investigate the relation to approaches via the Poynting vector \cite{Joulain2008, Chen2010}, nonlinear temperature profiles, as well as a deeper analysis of the occurring surface modes.

\begin{acknowledgments}
We thank D. Gelbwaser-Klimovsky, N. Graham and T. Emig for discussions. We acknowledge the KITP at UCSB where this work was initiated during the program Emerging Regimes and Implications of Quantum and Thermal Fluctuational Electrodynamics (supported by the National Science Foundation  Grants  PHY-1748958 and PHY-2309135). K.A. is supported by the Deutsche Forschungsgemeinschaft (DFG, German Research Foundation) through the Walter Benjamin fellowship (Project No. 453458207). M.K. acknowledges support from NSF grant DMR-2218849, and a Humboldt Research Award. 
\end{acknowledgments}





%
\end{document}